\magnification 1100
\hsize=6.0truein
\vsize=8.5truein
\baselineskip=16pt

\centerline {\bf A Theory of DDT in Unconfined Flames}
\bigskip
\centerline {Alexei M.\ Khokhlov,* Elaine S.\ Oran,** J.\ Craig Wheeler*}
\smallskip
\centerline {* Department of Astronomy}
\centerline {University of Texas at Austin}
\centerline {Austin, TX 78712}
\bigskip
\centerline {** Laboratory for Computational Physics}
\centerline {Naval Research Laboratory}
\centerline {Washington, DC 20375}

\vfill
\noindent
Corresponding Author:

Dr.\ Alexei M.\ Khokhlov

Department of Astronomy

University of Texas at Austin

Austin, TX 78712

phone: 512-471-3397 

fax: 512-471-6016

e-mail: ajk@astro.as.utexas.edu

\vfill

\centerline {Submitted to {\sl Combustion and Flame}, February 1995.}
\centerline {Revised version submitted December
1995.}
\centerline {Accepted March 1996.}

\vfill \eject

\centerline {\bf A Theory of DDT in Unconfined Flames}

\bigskip
\centerline {Alexei M.\ Khokhlov,* Elaine S.\ Oran,** J.\ Craig Wheeler*}

\smallskip\centerline {* Department of Astronomy}
\centerline {University of Texas at Austin}
\centerline {Austin, TX 78712}

\bigskip\centerline {** Laboratory for Computational Physics}
\centerline {Naval Research Laboratory}
\centerline {Washington, DC 20375}

\bigskip\centerline {\bf Abstract}

\bigskip\noindent
This paper outlines a theoretical approach for predicting the onset of
detonation in unconfined turbulent flames. Two basic assumuptions are made: 1)
the gradient mechanism is the inherent mechanism that leads to DDT in
unconfined conditions, and 2) the sole mechanism for preparing the gradient in
induction time is by turbulent mixing and local flame quenching. The criterion
for DDT is derived in terms of the one-dimensional detonation wave thickness,
the laminar flame speed, and the laminar flame thickness in the reactive gas. 
This approach gives  a lower-bound criterion for DDT for conditions where shock
preheating, wall effects, and interactions with obstacles are absent. Regions
in parameter space where unconfined DDT can and cannot occur are determined.

\bigskip
\noindent
\centerline {\bf 1. Introduction}
\medskip\noindent
The quantitative prediction of deflagration-to-detonation transition (DDT) in
energetic gases is one of the major unsolved problems in combustion and
detonation theory. Predicting the occurance of DDT has practical importance
because of its destructive potential. It is also an extremely interesting and
difficult scientific problem  because of the complex nonlinear interactions
among the different contributing physical processes, such as turbulence, shock
interactions, and energy release. An early description of experiments on DDT is
given by Brinkley and Lewis [1], who also describe Karlovitz's theory [2]. Much
of this theory has subsequently been experimentally confirmed and expanded upon
by Oppenheim and coworkers [3--5]. Excellent reviews that summarize our
understanding to date have been given by Lee and Moen [6] and, most recently,
by Sheppard and Lee [7]. Other useful summaries of mechanisms of DDT have been
given by Lewis and von Elbe [8] and Kuo [9].

Turbulence plays an important role in DDT. Several apparently different
mechanisms for the DDT in confined conditions have been described, each
including the turbulence of the flame and formation of shocks. On large scales,
turbulence deforms the flame front and increases its surface area. On small
scales, it broadens the flame front and causes mixing.  The result is an
extended turbulent ``flame brush'' in which a series of explosions occurs, one
of which finally leads to a detonation. Other routes to detonation  may include
an explosion in the boundary layer, or an explosion inside the region between
the leading shock and flame brush.

It is believed that, in most cases, the intrinsic mechanism triggering a 
detonation is the explosion of a nonuniformly preconditioned region of fuel in
which a spatial gradient of induction time has been created either by turbulent 
mixing, shock heating, or  both.  This gradient mechanism, first suggested by
Zeldovich and collaborators for nonuniform temperature  distributions  
[10,11], was subsequently found in photo-initiation experiments by Lee at al.\
[12], who called it the SWACER mechanism. This mechanism has since been studied
and described extensively (see, for example, [6,13--16]).  The mechanisms
for preconditioning the region, that is, the mechanism for preparing an
explosive mixture that has a gradient in induction times, may differ in
different situations. It may be created by a shock wave, turbulence,
photo-irradiation, intrinsic flame instabilities, rarefaction, or a combination
of several of these.

It appears to be very difficult to obtain DDT in unconfined conditions
[17--19]. This can be attributed to the geometrical effects of expansion: 
shocks which precede a deflagration might be weakened, or turbulence might be
damped too much by the expansion, and so  become unable to precondition the
mixture. Wagner and coworkers [17] report experiments in  which deflagrations
were forced to DDT by passing through screens of specified mesh sizes. The
screens created turbulence of the required scale and intensity.  These
experiments suggest that an unconfined  deflagration could make the transition
to detonation under the right  conditions.  This possibility has been suggested
for very large vapor clouds [6,20]. 

A related problem that has been studied experimentally is initiation of
detonations by turbulent jets [18,21--24]. In these experiments, a jet of hot
product gases in injected into an unburned, cold mixture. The turbulence
generated by the interaction of this jet and the background gas created  a
nonuniform, preconditioned region in which detonation may occur. For these
experiments, the effects of reflected shocks and interaction with walls is
minimal compared to DDT in tubes. Therefore, these experiments provide
important information on the critical size of the region capable of triggering
DDT.

We then ask the following question: What are the minimal requirements for DDT
in an idealized situation when all wall effects and incident shocks are
eliminated? If we can answer this question, we have a lower bound for DDT
conditions. Knowing the necessary conditions for unconfined DDT, we may then
draw conclusions about the relative importance of wall effects and shocks of
different strengths. One possible application of this theory could be to create
reproducible detonations in the shortest time and smallest space, as required
for pulse-detonation engines. Another application is to the theory of
supernovae. If DDT does occur in supernovae, as indicated by observations
[25,26], it would arise from an unconfined transition. Currently, there is no
quantitative theory explaining exactly how and when an unconfined transition
would occur.

In this paper, we  derive a theory for unconfined DDT.
That is, we address the situation where there are no external or reflected 
shocks, and no wall effects. 
We make two basic assumptions: 
\smallskip

\item{\it i.} The gradient mechanism  is the inherent mechanism that leads to
DDT in unconfined conditions, and

\item{\it ii.} The sole mechanism for preparing the gradient in induction
time is by turbulent mixing and local flame quenching. By this assumption, the
role of turbulence is to mix high-entropy products of burning and low-entropy
unreacted fuel. Such mixing creates gradients of temperature and concentration
which have opposite signs. Turbulence-generated shocks are ignored. 
\smallskip

Given these assumptions, there are two fundamental questions to address:  1)
What is the minimum size of a mixed region capable of generating a detonation, 
and  2) What level of turbulence is required to create this region?   We
address these two questions separately, and then  combine the answers to derive
the conditions for unconfined DDT.  Here we do not address the question how
these conditions may be produced, but give the scale and intensity of the
turbulence that is required. The derived criterion gives  {\sl lower bounds} on
conditions for DDT that  does not take into account many secondary effects that
may facilitate DDT. We then conclude with a discussion of the quantitative
importance of secondary effects.

\vfill\eject
\bigskip\centerline{\bf 2. Critical Size of the Preconditioned Region}

\bigskip\noindent
In this section we address the first of the two questions formulated in the
introduction. We consider the process of the initiation of detonation that
arises from the explosion of reactive gas with a nonuniform  distribution of
induction times. We assume that the nonuniformity  is a result of mixing of
high-entropy products and low-entropy unreacted fuel. We determine the minimum
size $L_c$ of a mixed region capable of triggering  a detonation. Whether and
how such a region can be created is a separate question that is studied in
Section 3. 

We can imagine a variety of regions of different shapes and degrees of mixing
created by turbulence. Here we consider the  simplest representative case of a
mixed region with a {\sl linear} one-dimensional distribution of products. In
the future, we plan to consider regions with different shapes, and thus explore
the influence of geometry. However, we do not believe geometrical
considerations will qualitatively  change our conclusions (Section 3.2).

\medskip
\centerline {\sl 2.1 Spontaneous Burning}
\smallskip
\noindent
To facilitate the discussion of a nonuniform explosion of a mixed region, it is
useful to discuss in general terms the idea of spontaneous burning.  This
concept was first introduced in Zeldovich and Kompaneetz [27]. Consider a
mixture with a nonuniform distribution of temperature $T(x)$ and chemical
composition $Y(x)$. The induction time  becomes a function of spatial
coordinate, $\tau(T(x),Y(x))$.  In the absence of any physical communication
between different  fluid elements, the explosion will start at a point of
minimum $\tau$, and then will spread spontaneously with a ``phase'' speed 
$$
   D_{sp} = \left(d\tau\over dx\right)^{-1}~,                       \eqno(1)
$$
which can have any value from zero to  infinity. A spontaneous reaction wave
does not require any physical agent in order to spread. Therefore, its  speed
is not limited by the speed of light. In reality, there is physical
communication between fuel elements. If the spontaneous  velocity is too small,
shocks and even heat conduction may cause faster flame propagation than that
prescribed by equation (1). 

Let $\delta t$ be the time during which the  bulk of chemical energy is 
released after the induction period is over, $\delta t << \tau$. We can define
the thickness of the spontaneous wave as  $\l_{sp} \simeq D_{sp} \delta t$. If
$D_{sp} \rightarrow \infty$, the thickness of the wave also goes to $\infty$.
This corresponds to a constant volume explosion. If $D_{sp}$ is comparable to
the speed of a detonation, on the other hand, its thickness is also comparable
to the thickness of a detonation wave. In this latter case, $\delta_{sp}$ may
become much less than the size of the system under consideration. Then the
spontaneous  wave may be viewed as discontinuity which obeys the Hugoniot
relations for a discontinuity with  energy release.

On a pressure -- specific volume plane, spontaneous burning is represented
by points located on a detonation adiabat. This is shown on Figure 1a, where
the regimes of spontaneous burning occupy the part of the detonation adiabat
from point I to point CJ. The position of the spontaneous regime on the
adiabat is determined by the intersection of the  Rayleigh line $dP/dV = -
(D_{sp}/V_0)^2$ with the detonation adiabat. The regime I corresponds to an
infinite spontaneous velocity when all matter burns simultaneously due to
uniform preconditioning. The point CJ corresponds to the minimum possible
velocity of a steady spontaneous wave, and is equal to the Chapman-Jouguet
velocity of a steady detonation. The same part of the detonation adiabat, I --
CJ, is occupied by weak detonations. The difference between spontaneous waves
and detonations, is that there is no shock wave present inside a spontaneous
wave. The structure of a Chapman-Jouget detonation and a spontaneous wave of
the same strength is shown schematically in Figure~1b.

In a detonation, the material is first shocked (point S in Figure 1), and then
expands towards the CJ point along the S -- CJ line. In the corresponding
spontaneous wave, the material is continuously compressed along the O -- SP
line until it reaches the CJ point (or some other point SP). The pressure,
density, and velocity in a spontaneous wave become larger than those of a
constant volume explosion  (point I) because burning does not proceed
simultaneously! There exists a pressure gradient inside the wave pointing
opposite to its direction of propagation, since at any instant the wave
consists of fluid elements with different amounts  of released energy.  As a
result, a fluid element passing through the wave is compressed and accelerated
by this gradient. The slower the wave moves, the longer is the time spent
inside the wave, and the greater are the pressure, density and velocity jumps
across the wave. The principle of causality is not violated in the spontaneous
wave, as explained in [28]. Although the speed of the spontaneous wave is a
phase speed, it is a real supersonic wave of burning which looks like a
detonation in terms of the hydrodynamic parameters of burned material. 

We have discussed the situation where the spontaneous wave speed is greater or
equal to the CJ detonation velocity, $D_{CJ}$.  Suppose the gradient in
induction time is such that $ D_{sp} $ is initially greater than $D_{CJ}$, but
then it decreases so that it becomes less than $D_{CJ}$. In this case, when the
spontaneous wave crosses the CJ threshold, the burned material
immediately behind the wave, which moves with the local sound speed relative to
the wave, will tend to overcome the wave and produce a shock.  Consider an
intermediate regime with such a shock, O -- O' -- S' -- CJ, shown in Figure~1. 
First, material burns in a spontaneous wave from O -- O', then it is shocked to
point S', and then returns to the CJ point.  The transition from the
spontaneous wave to a CJ detonation may then proceed through a sequence of such
regimes, with increasing shock strengths.

The description given in the last paragraph is a quasi-steady  picture that is
applicable only if the spontaneous wave velocity changes slowly enough. If the
spontaneous wave velocity changes too fast, that is, the gradient is too steep,
the shock and reaction will separate, and the CJ detonation will not form. In
the process of transition from spontaneous wave to CJ detonation, the
spontaneous velocity must change slowly enough so that the shock and reaction
do not separate.  This means that the nonuniform region must be large enough so
that this separation does not occur, and this, in turn, gives a criterion for
unconfined DDT.

\bigskip
\centerline{\it 2.2 Formulation of the Problem}

\medskip\noindent
Consider an idealized one-dimensional system with the equation of state
$P=(\gamma-1)E_t$ and  $T=P/\rho$, where $P$, $T$, $\rho$ and $E_t$ are the
pressure, temperature, mass density and thermal energy density, respectively. 
The chemical reaction is described by a first-order Arrhenius expression,
        $$ {dY \over dt} = -Y \exp\left(-{Q\over T}\right)~, \eqno(2) $$ 
where $Q$ is the activation energy, and the chemical variable $Y$ ranges 
from $Y=1$ for pure reactants  to $Y=0$ for total products. 
Units of distance and time are such that the pre-exponential factor in equation
(1) is unity, and the gas constant is $R=1$.
Planar geometry  is assumed. The system obeys the Euler equations,
$$
\eqalign{
   &  {\partial \rho \over \partial t} ~+~
      {\partial \over \partial x}  \left( \rho U \right)~=~0~,         \cr
   &  {\partial \rho U \over \partial t} ~+~
      {\partial \over \partial x} \left( \rho U^2 + P \right)~=~0~,        \cr
   &  {\partial E \over \partial t} ~+~
   {\partial \over \partial x} \left( \, \left( E+P \right) U \, \right)~=~0~,
                                                       \cr} \eqno(3)
$$
where $U$ is the fluid velocity, $E = E_t + \rho U^2/2 - \rho qY$
is the total energy density including chemical energy, and $q$ is the
total energy release per unit mass.

The initial temperature and density of the fuel are $T_0$ and $\rho_0$. The
products of isobaric burning, an approximation to burning in a laminar flame, 
have a temperature $T_1 = T_0 + q  \left( {\gamma -1} \over \gamma \right)$. By
our assumption, we consider a nonuniform region created by {\sl
mixing} the products of isobaric burning and fresh fuel, such that there is  a
linear spatial distribution of reactants $Y(x)$ and temperature $T(x)$,
$$
\eqalign{
& Y(x) = \cases{  x/L ,& $0 \le x \le L$  \cr
                1   ,& $x > L$ \cr}               \cr
& T(x) = T_1 - ( T_1 - T_0 ) \, Y(x)~,               \cr}       \eqno(4)
$$
where $L$ is the size of the mixed region. Initially, the velocity of the
material is zero, and the pressure $P_0$ is constant everywhere.  The boundary
conditions at $x=0$ are reflecting walls (symmetry conditions).

The system is prepared in an initial state and then evolves in time,
first until  ignition takes place, then to the formation (or failure) of 
detonation, and then to the time when the generated detonation or shock leaves
the computational domain. The cases considered are listed in Table 1. Parameters
for the standard case H1 with $P_0=1$ and $T_0=1$ are chosen to approximate a
detonation in a stoichiometric hydrogen and oxygen mixture at pressure of 1 atm
and temperature of 293 K [29,30].  The extra cases, H2 ($T_0=2$) and H3
($T_0=3$), are considered to study the sensitivity of the detonation formation
to the  initial temperature of the fuel. 

The system of equations (2) and (3) is integrated numerically using a
one-dimensional version of a time-dependent, compressible fluid code based on
the Piecewise Parabolic Method (PPM) [31,32]. PPM is a second  order
Godunov-like method which incorporates a Riemann solver to describe shock
waves accurately. Shocks are typically spread on one or two computational cells
wide. A piecewise parabolic advection algorithm advects sharp shockless
features, such as density and composition discontinuities or gradients, without
diffusing them excessively or changing their shape. Contact discontinuities are
typically kept two or three cells wide. Details of the implementation are given
in [33,34]. The chemical reaction is coupled to fluid dynamics by time-step
splitting.  The kinetic equation~(2) is integrated together with the equation
of energy conservation using adjustable substeps to keep the accuracy better
than 1\%. The grid spacing is selected so that there are at least 10 cells
within  a detonation wave reaction zone. The convergence of numerical solutions
was thoroughly tested by varying the number of computational cells from 1024 to
as many as 65536 in some cases.

\bigskip\centerline{\it 2.3 Detonation Formation Inside the Mixed Region}

\medskip\noindent
The induction time $\tau$ as a function of temperature $T$
and fuel fraction $Y$ can be expressed as
$$ 
\tau(T,Y) \simeq {1\over (\gamma -1)qY}\left(T^2\over Q\right) \, 
           \exp\left(Q\over T\right)~                           \eqno(5)
$$
using the Frank-Kamenetskii approximation [35], valid  when $T/Q << 1$, and
assuming the induction takes place at constant volume.  The derivatives of
$\tau$ with respect to $T$ and $Y$ are
$$ 
{\partial\tau\over\partial T} \simeq - {\tau Q \over T^2}~,~~~~~
{\partial\tau\over\partial Y} \simeq - {\tau \over Y}~. 
                                                                \eqno(6) 
$$
For the mixture considered here, the values of $T$ and $Y$ are  related by
equation (4). The function $\tau (T,Y(T))$ then has a minimum at $T_m$, found
by solving ${d\tau / dT} = {\partial \tau / \partial T} +  { ({\partial \tau /
\partial Y}) {(dY / d T)} } = 0$, so that
$$
 T_m^2 + Q T_m - QT_1 = 0~. \eqno(7)
$$
This gives
$$ 
\eqalign{ &T_m ~=~ Q {\sqrt{1 + 4{T_1\over Q}} - 1\over 2} 
               ~\simeq~ T_1 \left(1-{T_1\over 2Q}\right)~,      \cr
          &x_m ~=~ L\, \left(T_1-T_m\over T_1-T_0\right)
                ~\simeq~{LT_1\over 2Q}~,    \cr}
                                                                  \eqno(8)
$$
for $T_0 << T_1$.
The point $x_m$ is the first to ignite. From this point,
a spontaneous reaction wave propagates with the speed
$$
  D_{sp} ~=~ \left(d\tau\over dx\right)^{-1}
         ~=~ {T^2 (T-T_1) \over (T_1-T_0)(T^2 + QT -QT_1)}\, {L\over\tau}~.
                                                         \eqno(9)
$$
By virtue of equation (7), the speed of the reaction wave is infinite at point 
$x_m$. Thus, the reaction wave initially propagates supersonically, as
described in Section~2.1 We are interested in the propagation of the wave to the
right,  $x\rightarrow L$, where the energy released by the wave increases. 
The velocity of the wave decreases towards larger $x$, and becomes equal to the
local sound speed at some point $x_s$ determined by
$$
 D_{sp}(x_s) = D_{CJ}~.         \eqno(10)
$$
At this point, a pressure wave forms which runs into the mixture ahead of the
decelerating reaction wave. Whether this pressure wave is strong enough to
accelerate burning and to  evolve into a detonation wave depends on the length
$L$  of the mixed region.

There are two processes involved in the transformation of the pressure wave
into a detonation. First, the pressure wave must steepen into a shock.  This
shock must accelerate burning so that a shock--reaction complex forms. Second,
the shock--reaction complex must survive the propagation down the  temperature
gradient.  We denote as $L_s$ the first critical length of the mixed region
such that for $L < L_s$ the  shock--reaction complex does not form. For $L >
L_s$, the shock--reaction complex successfully forms within the mixed region.
We denote as $L_d$ the second critical length of the mixed region such that for
$L > L_d$, the shock--reaction complex  survives and passes as a detonation
into the cold fuel. For $L_s < L < L_d$, the shock--reaction  complex dies
inside the mixed  region.

Values of $L_s$ and $L_d$ were determined by the numerical simulations 
described in Section 2.2 by performing a series of simulations in which the
size $L$ of the mixed region was varied. Figures 2--5 show results
of simulations for H1 for four different choices of $L$. Each figure shows the
evolution with time of the pressure, velocity, temperature, and reactant
concentration. Figure 2 shows the results for the smallest mixed region, 
$L=30 x_d$, where $x_d$ is the half-reaction width of a CJ detonation. This
region is so small that the quasi-steady spontaneous wave cannot form. The
pressure wave is too weak to form a shock--reaction complex.  The pressure wave
generated by the spontaneous burning steepens into a shock outside of the mixed
region. 

For $L = 300 x_d$, shown in Figure 3, a shock wave forms at the point predicted
by equation (10), and the complex forms. The  simulations show that the
shock-reaction complex is far from a steady CJ detonation and cannot be
described as a small quasi-steady deviation from the CJ state. The peak
pressure is  at least a factor of two less than the von Neumann pressure for
the equivalent CJ detonation at the local condition. Figure 3 shows that, soon
after the  complex is formed, the reaction zone and shock  wave separate, and
only a shock wave leaves the mixed region. This is because the shock--reaction
complex, after being formed, must propagate through the mixture with
continuously decreasing temperature. The temperature  gradient  causes rapid
decrease in the postshock temperature, and, consequently, rapid growth of the
induction time in the postshock material.

Figures 4, $L = 500 x_d < L_d$, shows a case similar to Figure 3, but the
shock--reaction complex decouples close to the end of the mixed region.  In
Figure 5, $L= 960 x_d> L_d$, the complex  transforms into a detonation, and
passes into the cold unmixed fuel. The critical condition for the initiation of
detonation in mixed fuel and products is that the shock--reaction zone complex
survives its propagation through the temperature gradient. The critical
lengths, $L_d$, of a region capable of triggering a detonation, as determined
by such simulations,  are presented in Table 1 for cases H1 -- H3. 

The value of the critical length $L_d$ is sensitive to initial temperature
$T_0$. An increase of $T_0$ facilitates the initiation of detonation. Cases H2
and H3 in Table 1 show that $L_d$ decreases by a factor of six  if the initial
temperature is tripled. This can be explained if the criterion for the
detonation  formation is not the creation, but rather the survival of the
shock--reaction  complex. For higher initial temperature,  the postshock
induction time is less sensitive to variations of background conditions (see
equation 6), and so it is easier for the shock--reaction complex to adjust to
changing conditions.

\vfill \eject

\medskip
\centerline {\sl 2.4 Relation to Jet Initiation Experiments}
\smallskip

\noindent
One possible check on the theory described above for determining $L_d$ is to
compare the predictions of Section 2.3 with the results of turbulent
jet-initiation experiments [18,21--24]. In these experiments, a jet of hot
product gases in injected into an unburned, cold mixture. The jet can be 
characterized by the size of the orifice, $d$, through which hot products are
injected.  The turbulence caused by the interaction of this jet and the
background gas creates a nonuniform, preconditioned region in which detonation
may occur. The largest scale of the turbulence and the size of the mixed region
are also characterized by $d$. For these experiments, the effects of reflected
shocks and interaction with walls is believed to be small. The velocity of the
jet is approximately sonic with respect to the unburned background material,
Thus the strengths of the shocks formed by the exiting jet resulted in
overpressures in the unburned gas of about a factor of two or less. The
temperature increase was small. Ignition occurred in the jet and seemed to
be unaffected by wall interactions.

Depending on $d$, two distinct ignition regimes were found. For small $d$,
deflagrations were ignited at many points inside the mixed region. There was no
transition to detonation. For larger $d$, there was an explosion in the mixed
region that led to detonation. From these experiments, the minimum value of $d$
for which DDT occurred was $d > 10-20 l_c$, where $l_c$ is the detonation cell
size. 

The half-reaction zone length $x_d$, in terms of which we derived our estimates
of $L_d$, is a theoretical parameter. What is measured in experiments is a
detonation cell size $l_c$. In order to estimate $l_c$ for the case H1, we use
the results of two-dimensional simulations of detonation cell structure for
conditions similar to H1 [30]. Scaling the results of these simulations to
nondimensional units, we find $l_c \simeq 27 x_d$, where we have taken $l_c$ to
be the height of a detonation cell.  That is, the critical size of the mixed
region in case H1 is $L_d \simeq 36 l_c$.  Thus the theoretical estimate of
$L_d$  is in qualitative agreement with experiments. The somewhat larger
theoretical value, $36 l_c$ compared to $10-20  l_c$,  could be the result  of
the simplifying assumptions (one-step kinetics, neglect of multi-dimensional
effects) made in this paper. 

There have been other efforts to relate $L_d$ to $l_c$. For example, Knystautus
et al.\ [18] found that  $L_d \simeq 13 l_c$ based on the analogy between DDT
and direction initiation of detonation by an energy source. Dorofeev et al.\
[21] report $\L_d > 7 l_c$ based on their computations.

\vfill \eject
\bigskip\centerline {\bf 3. Critical Turbulent Velocity for DDT}

\bigskip\centerline {\it 3.1 Preconditioning by Turbulence}

\medskip\noindent
The discussion in the previous section established that the size of the region
required to trigger a detonation is large compared to the one-dimensional 
detonation thickness,  $L_d\simeq 10^3 x_d$ for case  H1.  Now the  question is
how to create this region. In an unconfined space, turbulence is the only
mechanism available. The  turbulence in the region of size $L_d$  must be
strong enough to create microscopic mixing in this region. Turbulence on large 
scales must be intense enough to pack individual laminar flame sheets close 
together. Turbulence on small scales must be strong  enough to broaden and
destroy individual flame sheets so that the products and fuel can mix to form
a microscopic region with a gradient of induction times. 

There are generally two regimes of turbulent flames we need to consider. The
first is a regime of multiple flame sheets, in which the flame is irregular
on large scales but laminar on small scales. The second is the distributed
burning regime, in which the turbulence is so strong that it modifies the
laminar flame structure (See, for example [36,37]. The transition between the
multiple flame sheet and distributed burning regimes represents the condition
where the creation of the large-scale nonuniform distribution of induction times
becomes possible. The flame will be affected by the turbulence on scales 
$\lambda \ge \lambda_G$, on which the turbulent velocity is greater than or
equal to the laminar flame speed, $S_l$. Here $\lambda_G$ is the Gibson scale
defined by the condition
$$ U (\lambda_G) = S_l \eqno (11) $$
where $U (\lambda)$ is the turbulent velocity on the scale $\lambda$. The
transition between the two turbulent regimes happens approximately when
$\lambda_G$ approaches the thickness of the laminar flame $x_l$ [36]. This
estimate is approximate and does not account for the effects of viscosity,
which becomes important when $\lambda_G$ approaches the viscous microscale
$\lambda_K$. The viscosity destroys turbulent eddies of size $x_l$. Poinsot et
al.\ [38] have shown theoretically that because of this effect, eddies larger
than $\lambda_G$ with velocitiy greater than $S_l$ are needed to quench the
flame. This has been substantiated by the experimental work by Roberts and
Driscoll [39] who showed that eddies a factor of four larger are required.

Consider, for simplicity, a Kolmogorov cascade inside the turbulent flame brush
such that on the scale $\lambda$, the turbulent velocity is
$$ 
U_\lambda \simeq U_{\cal L} \left( \lambda \over {\cal L} \right)^{1/3}~,
                                                               \eqno (12)
$$
where $\cal L$ is the driving scale of the turbulence, which could be
approximately equal to or larger than the size of the turbulent flame 
brush, and
$U_{\cal L}$ is the turbulent velocity on this scale. In this case, the Gibson
scale $\lambda_G$ becomes
 $$ \lambda_G \simeq \left( S_l \over U_{\cal L} \right)^3 \, {\cal L}~. 
                                         \eqno (13) $$ 
The condition $\lambda_G = x_l$ now can be used to define the intensity of the
turbulent motions needed for DDT,

$$
   U_{\cal L} = K\ S_l \left( {\cal L} \over x_l \right)^{1/3}~,  \eqno (14)
$$ 
where we introduced a coefficient $K\simeq 1$ which describes the ability of
the laminar flame to survive stretching and folding caused by turbulence on
scales of the order of $x_l$. Once the condition of equation (14) is reached
for ${\cal L} \geq L_d$, DDT can occur by the gradient mechanism.

For a typical flame, the thickness of the laminar flame $x_l$ is  approximately
an order of magnitude less than $x_d$. That is, $L_d \simeq 10^4 x_l$.  For a
flame with $L_d/x_l = 10^4$,  the intensity of turbulent motions required for
DDT on the scale of $L_d$ must be about $U_{L_d}
\simeq 20 S_l$, as follows from equation (14). For example, consider an
equimolar acetylene-oxygen flame  with a laminar flame speed of 5 m/s [40].
>From equation (12), the critical intensity of turbulent motions is
approximately $U_{\cal L} \simeq 100$~m/s, The critical turbulent velocity could
be considerably less in confined conditions because of the presence of shocks. 

In unconfined situations, there are two possible sources of turbulence,   the
Landau-Darrieus (L-D) instability and the Rayleigh-Taylor (R-T) instability.
The L-D instability is an intrinsic hydrodynamic flame instability that does
not require  external acceleration. The intensity of the L-D induced turbulent 
motions is unlikely to be much larger than $S_l$ because of nonlinear
stabilization effects [41]. The L-D instability is thus not likely to produce
the level of turbulence required for DDT in any  reasonable conditions. The
characteristic turbulent velocity associated with the R-T instability on scale
$L$  is of the order of $\simeq \sqrt{gL} \simeq 3 \sqrt{L}$ m/s for $L$  in
meters. The level of turbulence required for DDT can thus be achieved only on
scales of $\sim 100$ m.  This could explain why DDT in unconfined  situations 
is so rarely observed. To obtain DDT in the laboratory, we need  some other way
of inducing much higher turbulent intensities.

\bigskip\centerline{\it 3.2 Secondary Effects}

\medskip\noindent
When a region smaller than $L_d$ ignites, it can still generate a substantial
shock. The dependence of the maximum shock pressure on $L$ found from the
simulations is shown in Figure 6. The shock strength is high for $L$ larger
than, say,
$0.5L_d$, but rapidly decreases for smaller $L$. There are two possible effects
these shocks may produce, one related to the temperature increase and another
to vorticity created by the shocks.

The shock may raise the temperature in a region of the mixture that is  about to
explode, and this may facilitate the survival of the shock--reaction complex.
Table~1 shows that the increase of the initial temperature from $T_0=1$  to
$T_0=2$ decreases $L_d$ by a  factor of four. The increase of the initial
temperature by a factor of two requires, however, a shock strength $P_s/P_0
\simeq 8$. This shock strength  can be provided only by explosions of regions
of size $L > 0.5 L_d$ (see Figure 6). That is, this effect may slightly
decrease the  one-dimensional estimate of $L_d$, but is not likely to change it
drastically.

Another effect of a failed initiation on the surrounding material might be the
baroclinic generation of additional vorticity [1,42]. Such a secondary source
of turbulence reduces the amount of turbulence that must be generated by the
primary sources.  The turbulent velocity induced by this mechanism may be of
order of the postshock velocity. This source of secondary vorticity may be very
important in facilitating DDT, but only when the conditions are already close to
critical.  The amount of secondary vorticity will rapidly decrease with 
decreasing $L$. We conclude that our estimate of $L_d$ may  decrease by a 
factor of about two, but will not change drastically if the baroclinic 
mechanism is taken into account.

The major uncertainty in the estimation of the required turbulent velocity 
comes from our lack of exact knowledge of flame behavior on scales $\sim x_l$
in the turbulent velocity field. The standard definition of the Gibson  scale
as the scale at which the turbulent velocity is equal to the laminar flame
speed, $U_{\lambda_G} = S_l$, and the assumption that microscopic mixing 
begins when $\lambda_G = x_l$, gives $K=1$ in equation (14). As mentioned above,
recent work by Poinsot et al.\ [38] and Roberts and Driscoll [39] suggest $K
> 1$. There is also some evidence from numerical simulations of  turbulent
flames that this coefficient might be $K\simeq 3-5$, which would increase the
critical turbulent velocity accordingly [33]. This must be studied in future
numerical simulations and experiments.

The same kind of mixing and flame quenching must also take place in the flame 
brush of a  turbulent deflagration in a tube in order to have DDT in a confined
situation. Although shock preconditioning definitely plays an important role in
confined situations, there should be a  qualitative similarity between
triggering detonation by the explosion in the  middle of the brush and DDT in
unconfined conditions.  Carefully planned experiments on DDT in tubes with
quantitative characterization of the the turbulent velocity field prior to the
explosion in the brush  might be used to shed light on the exact value of
coefficient~$K$.

\bigskip\centerline {\bf 4. Conclusions}
\medskip\noindent
There are two key elements to the theory 
presented above for unconfined DDT:
\smallskip

\item{1.} {\it The size of the region $L_d$ that can trigger DDT in a mixture
of hot burning product and fuel}.  We estimate that $L_d \sim 10^3 x_d$, where
$x_d$ is the thickness of the one-dimensional reaction zone of the
Chapman-Jouguet detonation, or $ L_d \simeq 36~l_c$, where $l_c$ is the 
detonation cell size, or $L_d \simeq 10^4 x_l$, where $x_l$ is the laminar
flame thickness. This implies that large-scale mixing is required to
precondition the region.

\item{2.} {\it The intensity of turbulent motions required for the  region of
size $L_d$  to undergo DDT.} This is estimated from the requirements that the
Gibson scale inside this region be  comparable to  or less than the thickness
of the laminar flame (equation 14).  This requires the speed of the  turbulent
flame brush to be $\sim 10^2$ times faster than the laminar flame.

\smallskip
\noindent
The criterion of DDT in unconfined flames given here can be formulated in terms
of the following three  parameters of a reactive gas:  the one-dimensional
thickness of a CJ detonation, $x_d$, the velocity $S_l$, and the thickness $x_l$
of the laminar flame. The critical size of the mixed region $L_d$ can be 
directly related to $x_d$ (Section 2.3),  and the latter two parameters
determine the critical intensity of turbulence in the mixed region  required
for triggering DDT (Section 3.1).

The high turbulent velocity required for unconfined DDT is extremely difficult 
to achieve by turbulence generated by the flame itself or by the 
Rayleigh-Taylor instability,  which explains why DDT in unconfined flames is so
hard to observe. The critical size of the region $L_d$ derived in this paper
is in agreement with the results of hot jet initiation experiments. The theory
may also be extended to confined DDT in  the cases when the explosion leading
to detonation takes place in the middle of a turbulent flame brush. 

\bigskip\leftline {\bf Acknowledgments}
\medskip\noindent
This work was sponsored by the National Science Foundation, Texas Advanced
Research Project, the Naval Research Laboratory through the Office of Naval
Research, and the NASA Astrophysical Theory Program.  The authors are grateful
to C.J.\ Sung and C.K.\ Law for the useful information on laminar flame speeds,
to Vadim Gamezo for his unpublished results of detonation cell simulations in
hydrogen-oxygen mixtures,  to Martin Sichel for the discussion of general
properties of DDT, and to James F.\  Driscoll for references and discussion of
the cutoff scales. The computations were performed at the
High-Performance Computing Facility  of the University of Texas.

\vfill\eject
\centerline{\bf Table 1}

\vskip 0.80 cm
\centerline{\bf Table -- Simulated Cases}

\vskip 2.0 cm
\settabs\+\noindent&******&******&*******&*****
                   &********&********&********&********
                   &********&************&**********\cr

\hrule
\smallskip
\+&Case &$T_0$ &$\gamma$ &$q$ &$Q$ &$T_1$ &$T_s$ &$T_d$ 
                                          &$x_d$ &$L_s/x_d$ &$L_d/x_d$ \cr

\smallskip
\hrule
\smallskip
\+& H1 & $1$ & $1.333$ & $24$ & $28.3$ & $7.0$ & $5.8$ & $10.3$
       & $51.2      $ & $\sim 2\times 10^2$ &$ 9.5\times 10^2$ \cr
\+& H2 & $2$ & $1.333$ & $24$ & $28.3$ & $8.0$ & $7.1$ & $11.4$
       & $32.3      $ & - &$ 3.3\times 10^2$ \cr
\+& H3 & $3$ & $1.333$ & $24$ & $28.3$ & $9.0$ & $8.3$ & $12.5$
       & $23.5      $ & - &$ \sim 2\times 10^2$ \cr

\smallskip
\hrule

\vskip 2.0 cm
\settabs\+\noindent&***********&       \cr

\+& $T_0$    &Initial fuel temperature.  \cr
\+& $\gamma$ &Adiabatic index.  \cr
\+& $q$      &Total chemical energy release.  \cr
\+& $Q$      &Activation energy.  \cr
\+& $T_1$    &Temperature of products of isobaric burning. \cr
\+& $T_s$    &Postshock temperature in a Chapman-Jouguet detonation. \cr
\+& $T_d$    &Temperature of Chapman-Jouguet detonation products. \cr
\+& $x_d$    &Half reaction zone length of Chapman-Jouguet detonation. \cr
\+& $L_s$    &Critical length for shock--burning synchronization. \cr
\+& $L_d$    &Critical length for detonation survival in cold fuel.\cr

\vfill\eject

\centerline{\bf References}

\bigskip\noindent
1. Brinkley, S.R, Jr.\ and Lewis, B., {\sl Seventh Symposium (International)
  on Combustion}, The Combustion Institute, Pittsburgh, pp. 807--811, 1959.

\smallskip\noindent
2. Karlovitz, B.,{\sl Selected Combustion Problems}, p.\ 176,
    London, Butterworths, 1954.

\smallskip\noindent
3. Urtiew, P. and Oppenheim, A.K., {\sl Proc.\ Roy.\ Soc.\ Lond.\ A}, 
   295:13--28 (1966).

\smallskip\noindent
4. Laderman, A.J., Urtiew, P.A. and Oppenheim, A.K., {\sl Ninth Symposium
   (International) on Combustion},  The Combustion Institute, Pittsburgh,
    p. 265, 1963.

\smallskip\noindent
5. Oppenheim, A.K., Laderman, A.J. and Urtiew, P.A., {\sl Combust.\ Flame},
   6:193--197 (1962).

\smallskip\noindent
6. Lee, J.H.S. and Moen, I.O., {\sl Prog.\ Energy Combust. Sci.}, 6:359--389
  (1980).

\smallskip\noindent
7. Shepherd, J.E. and Lee, J.H.S., {\sl Major Research Topics in Combustion},
   pp.\ 439--490, eds.\ M.Y.\ Hussaini, A.\ Kumar, and R.G.\ Voigt, 
   Springer, New York, 1992.

\smallskip\noindent
8. Lewis, B. and von Elbe, G., 
   {\sl Combustion, Flames, and Explosions of Gases},
    Academic, 1987. 3rd edition,  pp. 566-573.

\smallskip\noindent
9. Kuo, K.K., {\sl Principles of Combustion}, Wiley, New York, 1986.

\smallskip\noindent
10. Zel'dovich, Ya.B., Librovich, V.B., Makhviladze, G.M. and
     Sivashinsky, G.I.,
     in {\sl 2nd International Colloquium on Explosion and 
     Reacting Systems Gasdynamics}, 1969, 
    Aug. 24--29. Novosibirsk, p.10.

\smallskip\noindent
11. Zel'dovich, Ya.B., Librovich, V.B., Makhviladze, G.M. and
     Sivashinsky, G.I., 
    {\sl Astronaut. Acta} 15:313--321 (1970).

\smallskip\noindent
12. Lee, J.H.S.,  Knystautas, R. and Yoshikawa, N., {\sl Acta Astronaut.}
     5:971--982 (1978).

\smallskip\noindent
13. Yoshikawa, N., {\sl Coherent Shock Wave Amplification in Photochemical
    Initiation of Gaseous Detonation}, Ph.D.\ Thesis, Department of Mechanical
    Engineering, McGill University, 1980.

\smallskip\noindent
14. Zel'dovich, Ya.B., Gelfand, B.E., Tsyganov, S.A., Frolov, S.M.
     and Polenov, A.N., {\sl Progr.\ Astronaut.\ Aeronaut.} 
     114:99-123 (1988).

\smallskip\noindent
15. He, L. and  Clavin, P., {\sl Twenty-Fourth Symposium (International)
   on Combustion}, The Combustion Institute, Pittsburgh,
    pp. 1861--1867, 1992.

\smallskip\noindent
16. Weber, H.-J.,   Mack, A. and Roth, P., {\sl Combust. Flame} 
    97:281--295 (1994).

\smallskip\noindent
17. Wagner, H.Gg., in {\sl Proc. Int. Specialist Conf.\ Fuel-Air 
    Explosions}, 
    U.\ Waterloo Press, 77--99 (1981).

\smallskip\noindent
18. Knystautas, R., Lee, J.H.S., Moen, I.O. and Wagner, H.Gg., 
    {\sl Proceedings of the 17th International Symposium on Combustion},
     The Combustion Institute, Pittsburgh, pp. 1235--1245, 1978.

\smallskip\noindent
19. Boni, A.A, Chapman, M., Cook, J.L. and Schneyer, J.P.,
 in {\sl Turbulent Combustion},  ed.\  L.\ Kennedy, pp.\ 379--405, (1978).

\smallskip\noindent
20. Moen, I.O., Lee, J.H.S., Hjertager, B.H., Fuhre, K. and Eckhoff, R.K.,
      {\sl Combust. Flame} 47:31--52, (1982).

\smallskip\noindent
21. Dorofeev., S.B., Bezmelnitsin, A.V., Sidorov, V.P., Yankin, J.G., and
Matsukov, I.D., {\sl Fourteenth International Colloquium on Dynamics of
Explosions and Reactive Systems}, University of Coimbra, Coimbra, 1993, Vol.\
2, pp.\ D2.4.1--D2.4.10.

\smallskip\noindent
22. Carnasciali, F., Lee, J.H.S., and Knystautas, R., {\sl Combust.\ Flame} 84:
170--180 (1991).

\smallskip\noindent
23. Moen, I.O., Bjerketvedt, D., and Jenssen, A., {\sl Combust.\ Flame} 61:
285--291 (1985).

\smallskip\noindent
24. Medvedev, S.P., Polenov, A.N., Khomik, S.V., and Gelfand, B.F., {\sl
Twenty-Fifth Symposium (Internqational) on Combustion}, The Combustion
Institute, Pittsburgh, pp.\ 73--78, 1994.

\smallskip\noindent
25. Khokhlov, A.M., {\sl Astronomy ans Astrophys.} 245:114 and 245:L25 (1991).

\smallskip\noindent
26. H\"oflich, P.A., Khokhlov, A.M. and  Wheeler, J.C. {\sl Astrophys.J.}
    444:831--847 (1994).

\smallskip\noindent
27. Zeldovich, Ya.B., and Kompaneets, S.A., {\sl Theory of Detonations}, New
York, Academic Press, 1960.

\smallskip\noindent
28. Khokhlov, A.M., {\sl Astron.\ Astrophys.} 246: 383--396 (1991).

\smallskip\noindent
29. Strehlow, R., Maurer, R.E. and Rajan, S., {\sl AIAA} 7:323--328 (1969). 

\smallskip\noindent
30. V. Gamezo (private communication).
 
\smallskip\noindent
31. Colella, P., and Woodward, P.R., {\sl J.\ Comp.\ Phys.} 54:174--201 (1984).

\smallskip \noindent 
32. Colella, P., and Glaz, H.M., {\sl J.\ Comp.\ Phys.} 59:264--289 (1985).

\smallskip\noindent
33. Khokhlov, A.M., {\sl Astrophys.J.} 449:695--713 (1995).

\smallskip\noindent
34. Khokhlov, A.M., Oran, E.S. and Wheeler, J.C.,
       {\sl Combust. Flame} (1996) 105:28--34.

\smallskip\noindent
35. Frank-Kamenetskii, D.A., {\sl Diffusion and Heat Transfer in Chemical 
    Kinetics}, 
   Nauka, Moscow (1967).

\smallskip\noindent
36. Peters, N., in {\sl Numerical Approaches to Combustion Modeling}, eds.\
E.S.\ Oran and J.P.\ Boris, AIAA, Washington, DC 1991, pp.\ 155-182.

\smallskip\noindent
37. Williams, F.A., in {\sl Mathematics of Combustion}, ed.\ J.D. Buckmaster,
Vol.\ 2, of {\sl Frontiers in Applied Mathematics}, SIAM, Philadelphia, pp.\
97--131 (1985).

\smallskip\noindent
38. Poinsot, T., Veynante, D., and Candel, S., {\sl J. Fluid Mech.} 228:
561--606 (1991).

\smallskip\noindent
39. Roberts, W.L., and Driscoll, J.F., {\sl Combust. Flame} 87: 245--256
(1991).

\smallskip\noindent
40. Sung, C.J. (private communication).

\smallskip\noindent
41. Zel'dovich, Ya.B., {\sl Journ. Applied Mech. Tech. Phys.} 7:68 (1966).

\smallskip\noindent
42. Picone, J.M., Oran, E.S., Boris, J.P. and  Young, T.R. Jr.,
 in {\sl Dynamics 
  of Shock Waves, Explosions, and Detonations},
  eds.\ J.R.\ Bowen, N.\ Manson,
  A.K.\ Oppenheim, and R.I.\ Soloukhin, 94,
  {\sl Prog.\ Astronaut.\ Aeronaut.},
   1985, pp.\ 429--488.

\vfill\eject

\centerline{\bf Figure captions}

\bigskip\noindent
Fig. 1 -- (a) Regimes of burning in the $P-V$ plane, ($V= 1 / \rho$). Solid
curve: detonation adiabat. Dashed curves: shock adabats. $O$, initial state of
cold fuel; $I$, state of products of constant volume explosion; $CJ$,
Chapman-Jouguet state; $S$, unburned post-shock state for CJ detonation; $S'$,
post-shock state for intermediate regime; $O'$, preshock state for
intermediate regime; $SP$, state of products of spontaneous burning. (b)
Schematic of pressure profile of a CJ detonation, spontaneous wave, and 
intermediate regime. For the spontaneous wave, the pressure changes smoothly
from $O$ to $CJ$, in contrast to the discontinuity $O$ -- $S$ in the
detonation. In the intermediate regime, pressure increases smoothly from $O$ to
$O'$, and then passes through the shock $O'$ -- $S'$.

\bigskip\noindent
Fig. 2 --  Pressure, velocity, temperature, and reactant concentration as a
function of distance at different times (marked by numbers in increasing order)
for case H1 (Table 1) with $L/x_d = 30$. Vertical arrow indicates where $D_sp
= D_{CJ}$. Values of $x_d$ are given in Table~1.

\bigskip\noindent
Fig. 3 --  Same as Fig.\ 1, but with $L/x_d = 300$.

\bigskip\noindent 
Fig. 4 --  Same as Fig.\ 1, but with $L/x_d = 500$. 

\bigskip\noindent 
Fig. 5 --  Same as Fig.\ 1, but with  $L/x_d = 960$.

\bigskip\noindent
Fig. 6 -- Maximum shock pressure $P_s$ generated during a nonuniform explosion
as a function of the length $L$ of a mixed region.

\vfill\eject\end